\documentclass[10pt,aps,pra,twocolumn,groupedaddress,showpacs]{revtex4-2}
\usepackage{bm}
\usepackage{graphicx}
\usepackage{amsmath}
\usepackage[usenames,dvipsnames]{xcolor}
\usepackage{dcolumn}
\newcommand*{\centt}[1]{\multicolumn{1}{c}{#1}}
\newcommand*{\cent}[1]{\multicolumn{1}{c}{$#1$}}
\newcolumntype{x}[1]{D{.}{.}{#1}}

\newcommand{\nr}{\vec{\nabla}_{\!R}}

\newcommand{\nel}{\vec{\nabla}_{\!\mathrm{el}}}

\newcommand{\SE}{Schr{\"o}dinger equation}
\newcommand{\suma}{\sum_{a}}
\newcommand{\me}{m_{\mathrm{e}}}
\newcommand{\na}{\vec{\nabla}_{\!a}}
\newcommand{\mun}{\mu_{\mathrm{n}}}
\newcommand{\Eel}{\mathcal{E}_{\mathrm{el}}}
\newcommand{\Ea}{\mathcal{E}_{\mathrm{a}}}

\newcommand{\EVS}[1]{\left\langle\phi\left|#1\right|\phi\right\rangle_{\!\mathrm{el}}}
\allowdisplaybreaks

\makeatletter
\DeclareRobustCommand{\cev}[1]{%
  \mathpalette\do@cev{#1}%
}
\newcommand{\do@cev}[2]{%
  \fix@cev{#1}{+}%
  \reflectbox{$\m@th#1\vec{\reflectbox{$\fix@cev{#1}{-}\m@th#1#2\fix@cev{#1}{+}$}}$}%
  \fix@cev{#1}{-}%
}
\newcommand{\fix@cev}[2]{%
  \ifx#1\displaystyle
    \mkern#23mu
  \else
    \ifx#1\textstyle
      \mkern#23mu
    \else
      \ifx#1\scriptstyle
        \mkern#22mu
      \else
        \mkern#22mu
      \fi
    \fi
  \fi
}
\makeatother
\newcommand{\rn}{\cev{\nabla}_{\!R}}
\newcommand{\nb}{\vec{\nabla}_{\!B}}
\newcommand{\bn}{\cev{\nabla}_{\!B}}

\begin{document}

\preprint{Version 0.1}

\title{Nuclear magnetic shielding in HD and HT}

\author{Mariusz Puchalski}
\affiliation{Faculty of Chemistry, Adam Mickiewicz University, Uniwersytetu Pozna{\'n}skiego 8, 61-614 Pozna{\'n}, Poland}

\author{Jacek Komasa}
\affiliation{Faculty of Chemistry, Adam Mickiewicz University, Uniwersytetu Pozna{\'n}skiego 8, 61-614 Pozna{\'n}, Poland}

\author{Anna Spyszkiewicz}
\affiliation{Faculty of Chemistry, Adam Mickiewicz University, Uniwersytetu Pozna{\'n}skiego 8, 61-614 Pozna{\'n}, Poland}

\author{Krzysztof Pachucki}
\affiliation{Faculty of Physics, University of Warsaw, Pasteura 5, 02-093 Warsaw, Poland}

\date{\today}

\begin{abstract}
We perform a calculation of the nuclear magnetic shielding in HD and HT molecules, with complete
and perturbative accounts for nuclear masses. From the difference in shielding, 
we obtain the deuteron and triton magnetic moments in agreement with the CODATA value,
with the accuracy limited only by nuclear magnetic resonance measurements.
Most importantly, our calculations indicate a potential for improved determination 
of nuclear magnetic moments.
\end{abstract}

\maketitle

\section{Introduction}
The most accurate determination of the nuclear magnetic moments is that of the proton, 
$\mu_p =   2.792\,847\,344\,62(82)\;\mu_N$ \cite{Schneider:17}. Magnetic moments of all other nuclei 
have been measured with less or much less accuracy. One of the reasons is the lack of a convenient 
reference system for which we accurately know the magnetic moment and which can be used for relative
measurement of nuclear magnetic resonance (NMR) frequencies.
$^3$He as a noble gas atom would be very convenient once its magnetic moment is accurately measured 
and the shielding calculated. In fact, the magnetic shielding in the $^3$He atom has very recently 
been calculated with the inclusion of leading relativistic and quantum electrodynamics effects to obtain 
$\sigma = 59.967\,029(23)\cdot10^{-6}$ \cite{Wehrli:21}. Also, the direct measurement 
of the helion ($^3$He nucleus) magnetic moment, like that of the proton, is being considered by the Heidelberg group 
\cite{schneider19}, which will allow $^3$He to be set as the ultimate reference system for NMR measurements. 
In the meantime, we plan to determine the helion magnetic moment from the relative measurements 
of NMR frequencies between the hydrogen molecule and the $^3$He atom \cite{Garbacz:12}. 

Let us now recall the definition of the magnetic shielding. When a molecule is placed in a homogeneous magnetic field $\vec B$, its nuclei experience 
the field that is shielded by the surrounding electrons $(1-{\sigma})\,\vec B$.
The magnitude of the shielding factor $\sigma$, typically of the order of $10^{-5}$, 
depends on the particular atomic and molecular system. 
Ramsey first considered this effect in~\cite{Ramsey:50} with the help of the nonrelativistic Hamiltonian with clamped nuclei 
in the external magnetic field. His result for the isotropic shielding factor $\sigma_0$ 
in the Born-Oppenheimer (BO) approximation is presented in Eq.~(\ref{shield_bo}). 
An immediate conclusion that can be drawn from this formula is that the shielding of the proton 
and deuteron (triton) in the HD (HT) molecule is the same. Clearly, one has to go beyond 
the BO approximation and include finite nuclear mass effects in the coupling 
to the external magnetic field to obtain the difference in the magnetic shielding.

 In the past there have been several attempts to calculate this shielding difference in HD (HT).  
Neronov and Barzakh in 1977 \cite{Neronov:77} derived the formula  for the shielding difference,
but they started with the incomplete Hamiltonian, i.e., their formula (4) does not include 
the nuclear spin-orbit interaction [see $g_A-1$ terms in Eq.~(\ref{eq:H}) below]. 
Later, calculations by Jaszu\'nski {\em et al.} \cite{Jaszunski:11} simulated nonadiabatic effects
by an artificial charge difference. Their result of $\delta\sigma({\rm HD}) = 9\cdot 10^{-9}$, 
although of the correct magnitude, is not well substantiated from the physical point of view, 
nor is it complete. In more recent calculations, Golubev and Shchepkin \cite{Golubev:14} 
used a more realistic treatment of nonadiabatic effects, 
but their result of $\delta\sigma({\rm HD}) = 9\cdot 10^{-9}$ was also incomplete.
In Ref.~\cite{Puchalski:15} we have derived the complete formula for the shielding difference 
and performed calculations with the result $\delta\sigma({\rm HD}) =20.20(2)\cdot 10^{-9}$; 
however, with some mistakes which are corrected here.

In this work, we calculate nuclear magnetic shielding in the HD and HT molecules and take advantage 
of the relative measurements of proton and deuteron (triton) NMR frequencies 
\cite{Garbacz:12,Neronov:77,Neronov:03,Neronov:11} 
\begin{equation}\label{eq:muAmuB}
\frac{\mu_A\,(1-\sigma_A)}{\mu_B\,(1-\sigma_B)} = \frac{f_A}{f_B}\,\frac{I_A}{I_B}\,.
\end{equation}
to determine deuteron and triton magnetic moments 
with the accuracy limited only by the experimental values of the NMR frequencies. 
For this, we present a rigorous derivation of the nuclear magnetic 
shielding constant. We obtain an exact formula that  applies for arbitrary nuclear masses, 
and we perform a so-called direct nonadiabatic (DNA) 
numerical calculation, treating the hydrogen molecule isotopologue as a four-body system. 
In addition, we derive the formula for the leading finite nuclear mass effects.
For this purpose, we employ the so-called nonadiabatic perturbation theory 
(NAPT)~\cite{PK:08,PK:09,Komasa:19} and point out a few mistakes in the former formula \cite{Puchalski:15}.
Numerical calculations show that these mistakes had only a minor influence on the nuclear shielding at the equilibrium 
distance, and our perturbative numerical results essentially agree with those obtained by us in Ref. \cite{Puchalski:15}.
Moreover, the obtained results using DNA agree with perturbative (NAPT) calculations. Therefore, we
confirm the recent CODATA~\citep{CODATA:18} values of the deuteron and triton magnetic moments which used
our previous results from Ref. \cite{Puchalski:15}.

\section{Theory of magnetic shielding accounting for the nuclear mass}  

The derivation of the nuclear magnetic shielding with full account for nuclear masses
closely follows that of Refs. \cite{Ramsey:50,Pachucki:10,Puchalski:15}. We start with the Hamiltonian 
for electrons and nuclei, which includes coupling to the external electromagnetic field
and all possible nuclear spin-orbit interactions, i.e.
\begin{eqnarray}
H &=& \sum_a\frac{\vec\pi_a^2}{2\,\me} + \frac{\vec\pi_A^2}{2\,m_A} + \frac{\vec\pi_B^2}{2\,m_B} + V
-\frac{e_A}{2\,m_A}\,g_A\,\vec I_A\vec B
\nonumber \\ &&
+\sum_b \frac{e_A\,e}{4\,\pi}\,\frac{\vec I_A}{2\,m_A}\cdot
\frac{\vec r_{Ab}}{r_{Ab}^3}\times
\biggl[g_A\,\frac{\vec\pi_b}{\me}-(g_A-1)\frac{\vec \pi_A}{m_A}\biggr]
\nonumber \\ &&
+\frac{e_A\,e_B}{4\,\pi}\,\frac{\vec I_A}{2\,m_A}\cdot
\frac{\vec r_{AB}}{r_{AB}^3}\times
\biggl[g_A\,\frac{\vec\pi_B}{m_B}-(g_A-1)\frac{\vec \pi_A}{m_A}\biggr],
\nonumber\\ \label{eq:H}
\end{eqnarray}
where we assumed $\hbar=c=1$, and where
$\vec \pi = \vec p-e\,\vec A$, $\vec A$ is an external magnetic vector potential, 
and $g_A$ is the $g$-factor of the nucleus $A$ related to the magnetic moment by
\begin{align}
\vec \mu_A = \frac{e_A\,g_A}{2\,m_A}\,\vec I_A.
\end{align}
To derive the formula for the shielding constant, including the finite nuclear mass
corrections, we perform a unitary transformation $\varphi$,
\begin{equation}
\tilde H = e^{-i\,\varphi}\,H\,e^{i\,\varphi}+\partial_t\varphi\,,
\end{equation}
which places the gauge origin at the moving nucleus $A$. 
We assume that the molecule is neutral and that the external magnetic field is homogeneous, so
\begin{eqnarray}
\varphi &=& \sum_a e\,\Bigl(x^i_a\,A^i + \frac{1}{2}\,x^i_a\,x^j_a\,A^i_{,j}\Bigr) 
\nonumber \\ &&
 + e_B\,\Bigl(x^i_B\,A^i + \frac{1}{2}\,x^i_B\,x^j_B\,A^i_{,j}\Bigr) ,
\end{eqnarray}
where $\vec A = \vec A(\vec r_A)$, and $\vec x_a = \vec r_a-\vec r_A$.
The transformed momenta are
\begin{eqnarray}
e^{-i\,\varphi}\,\pi^j_a\,e^{i\,\varphi} &=& p^j_a+\frac{e_a}{2}\,(\vec x_a\times\vec B)^j\,,\\
\nonumber\\[-1.7ex]
e^{-i\,\varphi}\,\pi^j_B\,e^{i\,\varphi} &=& p^j_B+\frac{e_B}{2}\,(\vec x_B\times\vec B)^j\,,\\
e^{-i\,\varphi}\,\pi^j_A\,e^{i\,\varphi} &=& p^j_A+\frac{e_A}{2}\,(\vec D\times\vec B)^j\,,
\end{eqnarray}
where $e_A\,\vec D = \sum_a e\,\vec x_a + e_B\,\vec x_B$ is the electric dipole moment operator.
We can now assume that the total momentum vanishes; thus, 
$\vec p_A = -\vec p_B -\sum_a\vec p_a$
and the independent position variables are $\vec x_a$ and $\vec x_B$.

The new Hamiltonian $\tilde H$ after this transformation
with $\vec p_{\rm el} = \sum_a \vec p_a$ and $\vec x_{\rm el} = \sum_a \vec x_a$  becomes
\begin{widetext}
\begin{align}
\tilde H =& \sum_a\frac{1}{2\,\me}\Bigl(\vec p_a+\frac{e}{2}\,\vec x_a\times\vec B\Bigr)^2
+\frac{1}{2\,m_B}\Bigl(\vec p_B+\frac{e_B}{2}\,\vec x_B\times\vec B\Bigr)^2
+\frac{1}{2\,m_A}\biggl(\vec p_B+\vec p_{\rm el}-\frac{e_A}{2}\,\vec D\times\vec B\biggr)^2 +V
\nonumber \\ &
-\frac{e_A\,g_A}{2\,m_A}\,\vec I_A\,\vec B
-\sum_a\frac{e_A\,e}{4\,\pi}\,\frac{\vec I_A}{2\,m_A}\cdot
\frac{\vec x_a}{x_a^3}\times\biggl[\frac{g_A}{\me}\,\biggl(
\vec p_a+\frac{e}{2}\,\vec x_a\times\vec B\biggr)
+\frac{(g_A-1)}{m_A}\,\biggl(\vec p_B+\vec p_{\rm el}-\frac{e_A}{2}\,\vec D\times\vec B\biggr)\biggr]
\nonumber \\ &
-\frac{e_A\,e_B}{4\,\pi}\,\frac{\vec I_A}{2\,m_A}\cdot
\frac{\vec x_B}{x_B^3}\times\biggl[\frac{g_A}{m_B}\,\biggl(
\vec p_B+\frac{e_B}{2}\,\vec x_B\times\vec B\biggr)
+\frac{(g_A-1)}{m_A}\,\biggl(\vec p_B+\vec p_{\rm el}-\frac{e_A}{2}\,\vec D\times\vec B\biggr)\biggr]\,.
\label{13}
\end{align}
Separating contributions that are linear in $\vec B$ and $\vec I_A$, one arrives at
  \begin{align}
    \tilde H =&\ H_0 + \vec H_B\cdot \vec B - \frac{e_A\,g_A}{2\,m_A}\,I_A^i\,\biggl(B^i+H_I^i  + H_{IB}^{ij}\,B^j\biggr) +\ldots
    \end{align}
    where $H_0$ is the nonrelativistic Hamiltonian of the hydrogen molecule and 
\begin{align}
 \vec H_B =&\
-\frac{e}{2\,\me}\sum_a\,\vec x_a\times\vec p_a
-\frac{e_B}{2\,m_B}\,\vec x_B\times\vec p_B
+\frac{e_A}{2\,m_A}\,\vec D\times(\vec p_B+\vec p_{\rm el}) \label{HB}\\
\vec H_I =&\ 
\sum_a\frac{e}{4\,\pi}\,\frac{\vec x_a}{x_a^3}\times\biggl[\frac{1}{\me}\,\vec p_a
+\frac{(g_A-1)}{g_A\,m_A}\,(\vec p_B+\vec p_{\rm el})\biggr]
+\frac{e_B}{4\,\pi}\,\frac{\vec x_B}{x_B^3}\times\biggl[\frac{1}{m_B}\,\vec p_B
+\frac{(g_A-1)}{g_A\,m_A}\,(\vec p_B+\vec p_{\rm el})\biggr] \label{HI}\\
H_{IB}^{ij}\,I_A^i\, B^j =&\
\sum_a\frac{e}{4\,\pi}\,\vec I_A\times
\frac{\vec x_a}{x_a^3}\,\biggl[\frac{e}{2\,\me}\,\vec x_a
-\frac{e_A}{2\,m_A}\,\frac{(g_A-1)}{g_A}\,\vec D\biggr]\times\vec B
+\frac{e_B}{4\,\pi}\,\vec I_A\times
\frac{\vec x_B}{x_B^3}\,\biggl[\frac{e_B}{2\,m_B}\,\vec x_B
  -\frac{e_A}{2\,m_A}\,\frac{(g_A-1)}{g_A}\,\vec D\biggr]\times\vec B\,.
  \end{align}
\end{widetext}
The coupling of the nuclear spin to the magnetic field is given by
\begin{align}
\delta E =&\ -\frac{e_A\,g_A}{2\,m_A}\biggl[\langle H_{IB}^{ij}\,I^i\,B^j\rangle +
2\,\langle \vec H_B\cdot\vec B\,\frac{1}{E_0-H_0}\,\vec H_I\cdot\vec I\rangle\biggr].
\end{align}
After averaging over orientations of the rotational angular momentum, $\delta E$ becomes
\begin{align}
\delta E &=
-\frac{e_A\,g_A}{2\,m_A}\,\frac{\vec I\cdot\vec B}{3}\,\biggl[\langle H_{IB}^{ii}\rangle +
  2\,\langle H_B^i\,\frac{1}{E_0-H_0}\,H_I^i\rangle\biggr],
  \end{align}
  where
  \begin{align}
  H^{ii}_{IB} &=
  -\sum_a\frac{e}{4\,\pi}\,\frac{\vec x_a}{x_a^3}\,\biggl[\frac{e}{m}\,\vec x_a
-\frac{e_A}{m_A}\,\frac{(g_A-1)}{g_A}\,\vec D\biggr]
\nonumber \\ &\quad\,
-\frac{e_B}{4\,\pi}\,\frac{\vec x_B}{x_B^3}\,\biggl[\frac{e_B}{m_B}\,\vec x_B
  -\frac{e_A}{m_A}\,\frac{(g_A-1)}{g_A}\,\vec D \biggr].
  \end{align}
Finally, the isotropic shielding constant is
  \begin{align}
\sigma =&\ -\frac{1}{3}\,\biggl[\langle H_{IB}^{ii}\rangle +
  2\,\langle H_B^i\,\frac{1}{E_0-H_0}\,H_I^i\rangle\biggr]. \label{isosh}
 \end{align}
This formula completely accounts for the nuclear masses and is employed in our numerical
calculations reported below. We note that evaluation of $\sigma$ according to~(\ref{isosh})
is not a straightforward task.
In the following sections, starting from the above result, we derive alternative simplified expressions 
for the leading finite nuclear mass effects using NAPT, and for the reader's convenience, 
we shall describe the NAPT matrix elements in Appendix~\ref{app:NAPT}.
 
\section{Nuclear magnetic shielding using NAPT}
Because the electron to nuclear mass ratio is small, it is customary to assume the BO approximation 
and represent the total wave function as a product of the electronic $\phi$ and nuclear $\chi$ functions
\begin{equation}
\psi(\vec r,\vec R) = \phi(\vec r;\,\vec R) \; \chi(\vec R). \label{BO}
\end{equation}
The electronic wave function $\phi$ depends parametrically on $\vec R$ and is the eigenstate of the clamped nuclei Hamiltonian $H_{\rm el}$ with an eigenvalue 
${\cal E}_{\rm el}(R)$, while $\chi(R)$ satisfies the nuclear equation with the Hamiltonian 
including potential ${\cal E}(R)$; for details, see Appendix~\ref{app:NAPT}. 
Analogously, physical quantities such as the magnetic shielding constant can be represented as 
an expectation value of the $R$-dependent quantity $\sigma(R)$ with the nuclear wave function $\chi$.
To obtain $\sigma(R)$ let us construct the general effective Hamiltonian that is the function 
of the internuclear distance and describes all the relevant interactions between the nuclear spin $\vec I_A$, the magnetic field $\vec B$, 
and the rotational angular momentum $\vec J$, namely
\begin{align}
H_{\rm eff}(\vec R) =&\ 
-\gamma_\mathrm{J}(R)\,\vec J\cdot\vec B - \gamma_\mathrm{I}(R)\,\vec I_A\cdot\vec J 
\nonumber \\ &\ - \gamma_A\,I_A^i\,B^j\,[\delta^{ij}-\sigma^{ij}(R)],
\end{align}
where
\begin{align}
\sigma^{ij}(R) =&\ 
\delta^{ij}\,\sigma(R) +  \sigma_T(R)\,(J^i\,J^j-\delta^{ij}\,\vec J^2/3) 
\end{align}
and where $\gamma_\mathrm{J}$ is the rotational magnetic moment, $\gamma_\mathrm{I}$ 
is the spin-rotational constant, and $\gamma_A = e_A\,g_A/(2\,m_A)$.
The isotropic shielding deduced from this Hamiltonian is
\begin{align}
\sigma &= \langle\chi|\sigma(R) |\chi\rangle\nonumber\\
\nonumber\\[-2ex]
&\quad + \frac{2\,J\,(J+1)}{3\,\gamma_A}\,\langle\chi|\gamma_\mathrm{J}(R)\frac{1}{(E-{\cal H})'}\gamma_\mathrm{I}(R)|\chi\rangle. \label{eq:sigmachi}
\end{align}
where $\cal{H}$ is defined in Eq. (\ref{Ead}).
The constants $\gamma_\mathrm{J}$, $\gamma_\mathrm{I}$, and $\gamma_A$ contain the inverse power
of nuclear masses, while the resolvent includes the sum over all vibrational excitations 
and is of the order of the inverse square root of the nuclear mass. Therefore, the latter term is smaller 
than the leading $m/\mun$ corrections ($\mun$ is the reduced nuclear mass) by the square root 
of the nuclear mass, which means it is negligible. However, $\gamma_{\rm I}(R)$ 
is the same for both nuclei at the  equilibrium distance---see formula (110) for the spin-rotation 
constant in Ref. \cite{Pachucki:10}. Therefore it cancels out in the shielding difference 
and the second term in Eq. (\ref{eq:sigmachi}) can safely be neglected. 

Considering now the first term in Eq. (\ref{eq:sigmachi}), the nuclear magnetic shielding $\sigma(R)$ 
is the sum of two terms
\begin{align}
\sigma(R) = \sigma_0(R)+\sigma_1(R).
\end{align}
Here $\sigma_0$ is the shielding in the BO approximation [neglecting all the terms in Eq. (\ref{isosh}) containing inverse powers of nuclear masses],
 \begin{align} 
\sigma_0(R) =&\ 
\frac{\alpha^2}{3}\biggl[\biggl\langle\sum_a\frac{1}{x_a}\biggr\rangle
\nonumber \\ &
+\biggl\langle\sum_a\,\vec x_a\times\vec p_a
\,\frac{1}{{\cal E}_\mathrm{el}-H_\mathrm{el}}
\sum_b \frac{\vec x_b}{x_b^3}\times \vec p_b \biggr\rangle\biggr] \label{shield_bo}
\end{align}
and $\sigma_1(R)$ is the first order in the electron-nuclear mass ratio correction.
We focus here only on terms that contribute to the difference in the nuclear shielding;
therefore, we take
\begin{align}\label{dsigmaR}
\sigma_1(R) = \frac{\alpha^2}{3}\,\frac{m_{\rm e}}{m_A}\,\biggl(\sigma_{\rm n}(R) + \sigma_A(R) + \frac{\sigma'_A(R)}{g_A}\biggr).
\end{align} 
The first correction $\sigma_{\rm n}$ is obtained by perturbing Eq. (\ref{shield_bo})
by the nuclear kinetic energy $H_{\rm n}$. As shown in Appendix~\ref{app:sigman}, 
$H_{\rm n}$ can be replaced  in matrix elements by $m/m_A\,\tilde H_{\rm n}$, with 
\begin{align}\label{eq:Hntilde}
\tilde H_{\rm n} \equiv
(\vec x_{\rm el}-\langle\vec x_\mathrm{el}\rangle)\,\nb[V-{\cal E}_{\rm el}]-\frac{\vec\nabla_{\rm el}^2}{2\,\me}\,,
\end{align}
plus some additional terms, which leads to
\begin{widetext}
\begin{align}
\sigma_{\rm n} =&\  \frac{\alpha^2}{3}\,\frac{m}{m_A}\,\biggl[
2\,\bigg\langle\sum_{b}\frac{1}{x_b}\,\frac{1}{{\cal E}_{\rm el}-H_{\rm el}}\,{\tilde H}_{\rm n}\bigg\rangle+
\Big\langle {\tilde H}_{\rm n}\,\frac{1}{{\cal E}_{\rm el}-H_{\rm el}}\,
\sum_a\vec x_a\times\vec p_a
\frac{1}{{\cal E}_{\rm el}-H_{\rm el}}
\sum_{b}\frac{\vec x_b\times\vec p_b}{x_b^3}\Big\rangle
\nonumber \\ &\ 
+\Big\langle\sum_a\vec x_a\times\vec p_a
\frac{1}{{\cal E}_{\rm el}-H_{\rm el}}\, {\tilde H}_{\rm n}
\frac{1}{{\cal E}_{\rm el}-H_{\rm el}}
\sum_{b}\frac{\vec x_b\times\vec p_b}{x_b^3}\Big\rangle
+\Big\langle\sum_a\vec x_a\times\vec p_a
\,\frac{1}{{\cal E}_{\rm el}-H_{\rm el}}
\sum_{b}\frac{\vec x_b\times\vec p_b}{x_b^3}
\frac{1}{{\cal E}_{\rm el}-H_{\rm el}} {\tilde H}_{\rm n}\Big\rangle
\nonumber \\ &\
+\biggl\langle\vec x_\mathrm{el}\times\vec P\,\frac{1}{{\cal E}_\mathrm{el}-H_\mathrm{el}}\sum_b \frac{\vec x_b}{x_b^3}\times \vec p_b\biggr\rangle
+\biggl\langle\sum_a\,\vec x_a\times\vec p_a\,\frac{1}{{\cal E}_\mathrm{el}-H_\mathrm{el}}\sum_b \frac{\vec x_b}{x_b^3}\times \vec P\biggr\rangle
\biggr], \label{eq:sigman}
\end{align}
\end{widetext}
where
\begin{align}
\vec x_{\rm el} =&\ \sum_a \vec x_a\,,\\
\vec \nabla_{\rm el} =&\ \sum_a \vec \nabla_a\,, \\
\vec P =&\ -\frac{i}{2}\,(\vec\nabla_R-\cev\nabla_R)\,. \label{29}
\end{align}
The latter, symmetrized form of the $R$-derivative comes from the careful analysis of matrix elements 
in the NAPT, as shown in  Appendix~\ref{app:NAPT}. The remaining corrections $\sigma_A$ and $\sigma'_A$ 
come from the explicit nuclear mass-dependent terms in Eq.~(\ref{isosh}). In these expressions,
the $1/m_B$ coefficient can be replaced by $1/\mun-1/m_A$, 
and the nuclear momentum $\vec p_B$ by $-\vec P$, which results in
\begin{align}
\label{sigmaA}
\sigma_{A} =&\
-\bigg\langle \frac{1}{R} +\vec K\cdot(\vec x_{\rm el}+\vec R)\bigg\rangle 
\nonumber\\ &\hspace{-7ex}
+\bigg\langle\sum_a\vec x_a\times\vec p_a\,\frac{1}{{\cal E}_{\rm el}-H_{\rm el}}
\biggl[\vec K\times\big(\vec p_{\rm el}-\vec P\big)
+\frac{\vec R}{R^3}\times\vec P\bigg] \bigg\rangle
\nonumber\\ & \hspace{-7ex}
+\bigg\langle\sum_a\frac{\vec x_a\times\vec p_a}{x_a^3}\frac{1}{{\cal E}_{\rm el}-H_{\rm el}}
\Bigl[\bigl(\vec x_{\rm el}+\vec R\bigr)\times \bigl(\vec P-\vec p_{\rm el}\bigr) + \vec R\times\vec P\Bigr]\bigg\rangle
\intertext{and}
\sigma'_{A} =&\ \bigl\langle\vec K\cdot(\vec x_{\rm el}+\vec R)\bigr\rangle \nonumber\\ &
+ \bigg\langle\sum_a\vec x_a\times\vec p_a\, \frac{1}{{\cal E}_{\rm el}-H_{\rm el}}\,\vec K\times(\vec P-\vec p_{\rm el})\bigg\rangle,\label{sigmaAp}
\end{align}
where
\begin{align}
\vec K = \sum_{a}\frac{\vec x_a}{x_a^3}+\frac{\vec R}{R^3}.
\end{align}
Formulas (\ref{eq:sigman}), (\ref{sigmaA}), and (\ref{sigmaAp}) almost coincide with our previous derivation in Ref. \cite{Puchalski:15}. 
The only difference is  in the presence of $\langle\vec x_\mathrm{el}\rangle = -\vec R$ 
in ${\tilde H}_{\rm n}$ of Eq.~(\ref{eq:Hntilde}) and in the symmetrization of the $\vec \nabla_R$ derivative. 
These changes affect the shape of the $\delta\sigma(R)$ curve, but close to equilibrium
the numerical values for HD and HT molecules are not essentially changed, as shown in the next section.

\section{Numerical calculations using explicitly correlated Gaussians}

Numerical evaluation of the shielding constant in the hydrogen molecule can be efficiently performed 
using explicitly correlated Gaussian (ECG) wave functions. In this work, we applied two independent 
ECG-based methods. The first one is the direct nonadiabatic (DNA) method, in which
all the particles of a molecule are treated on equal footing \cite{Puchalski:18},
and the second makes use of the nonadiabatic perturbation theory (NAPT) formalism. 
To a large extent, these two methods complement and verify each other. The most noteworthy feature
of the NAPT approach is the possibility of accurately determining all the rotational 
and vibrational energy levels simultaneously (for a given electronic
state). NAPT enables all the leading-order nonadiabatic effects to be accounted for  
in the Hamiltonian and wave function. In contrast, the DNA method fully accounts  (to all orders) 
for the finite nuclear mass effects and surpasses the NAPT approach in terms of accuracy, 
but each rovibrational level must be treated individually, making the method computationally expensive. 
Because we focus on properties of the rovibrational ground level, DNA is the method of choice.
On the other hand, the NAPT calculations gain significance at the stage of temperature averaging.

\subsection{Direct nonadiabatic approach}
\begin{table}[!htb]
 \caption{Convergence of the isotropic shielding constant $\sigma$ of Eq.~(\ref{isosh})
 and of the shielding difference $\delta\sigma(\mathrm{HX})=\sigma_x(\mathrm{HX})-\sigma_p(\mathrm{HX})$ 
 (in ppm) evaluated using the DNA method for the ground rovibrational level ($v=0,J=0$).}
 \label{DNAresultsHD}
 \begin{ruledtabular}
 \begin{tabular}{l x{3.10} x{3.11} x{2.10}}
        & \cent{\sigma_p(\mathrm{HD})} & \cent{\sigma_d(\mathrm{HD})} & \cent{\delta \sigma(\mathrm{HD})}\\ 
\hline
256     & 26.359\,335\,87   & 26.366\,390\,735   & 0.007\,054\,86  \\
384     & 26.352\,895\,60   & 26.372\,807\,241   & 0.019\,911\,64  \\  
512     & 26.352\,906\,53   & 26.372\,805\,619   & 0.019\,906\,14  \\
768     & 26.352\,902\,70   & 26.372\,804\,050   & 0.019\,901\,13 \\
1024    & 26.352\,901\,63   & 26.372\,801\,890   & 0.019\,900\,26  \\  
1536    & 26.352\,901\,31   & 26.372\,801\,914   & 0.019\,900\,61  \\
$\infty$& 26.352\,901\,1(3) & 26.372\,801\,9(10) & 0.019\,901(1)\\
\hline\rule{0pt}{2.5ex}
        & \cent{\sigma_p(\mathrm{HT})} & \cent{\sigma_t(\mathrm{HT})} & \cent{\delta \sigma(\mathrm{HT})}\\
\hline
256     & 26.366\,892\,93    & 26.392\,063\,42    & 0.025\,170\,5 \\
384     & 26.367\,539\,47    & 26.391\,387\,89    & 0.023\,848\,4 \\  
512     & 26.367\,542\,37    & 26.391\,389\,02    & 0.023\,846\,6 \\
768     & 26.367\,479\,23    & 26.391\,449\,91    & 0.023\,970\,6 \\
1024    & 26.367\,480\,06    & 26.391\,453\,89    & 0.023\,973\,8 \\  
1536    & 26.367\,480\,91    & 26.391\,455\,42    & 0.023\,974\,5 \\
$\infty$& 26.367\,481\,3(8)  & 26.391\,456\,1(15) & 0.023\,975(2) \\
\end{tabular}
\end{ruledtabular}
\end{table}

In the framework of the DNA method, the wave function for the ground-state hydrogen molecule was introduced 
in our previous papers \cite{Puchalski:18,Puchalski:19}, and here we recall only its most important
features. The total molecular wave function is represented in the form of a linear combination
\begin{eqnarray}\label{Psilc}
\Psi &=& \sum_i^N c_i \psi_i,\\
\psi_i &=& (1 \pm P_{0\leftrightarrow 1})\,(1+P_{2\leftrightarrow 3})\,\phi_i
\end{eqnarray}
of the four-particle Gaussian basis functions (called naECG)
\begin{equation}
\phi_{S} =r_{01}^n\, e^{-a_{01} r^2_{01}-a_{02} r^2_{02}-a_{03} r^2_{03}-a_{12} r^2_{12}-a_{13} r^2_{13}-a_{23} r^2_{23}}\,, \label{DNAECG} 
\end{equation} 
where the superscripts $0, 1$ and $2, 3$ correspond to nuclei and electrons, respectively. 
The nonlinear parameters $a_{ij}$ of each basis function were optimized variationally
with respect to the nonrelativistic energy. 
The integer powers $n$ of the internuclear coordinate $r_{01}$ were generated randomly 
from the log-normal distribution within the $0 - 80$ range. The final distribution of $n$
was obtained in an iterative refinement procedure by replacing basis functions of an insignificant
energy gain with new ones obtained from the updated distribution. For heterogeneous molecules, 
such as HD or HT, we distinguish functions that are symmetric and antisymmetric in the $0 \leftrightarrow 1$
exchange, and their share in the basis set was treated as another discrete optimization parameter.

The second-order matrix elements in Eq.~\eqref{isosh} involve the intermediate states 
of $P$-even ($P^e$) symmetry. The following basis functions represent such states:
\begin{eqnarray} \label{phiPe}
\vec \phi_{P^e} &=& (\vec r_{ab} \times \vec r_{cd})\,\phi_{S}
\end{eqnarray}
with arbitrary mapping of particle indices $0,\dots,3$ onto subscripts $a,\dots,d$.
The contribution of various variants of the angular prefactors was also determined 
in an iterative refinement process. 
The optimal shares of $0 \leftrightarrow 1$ symmetries and functions~(\ref{phiPe}) in the whole
basis set turned out to be crucial in obtaining highly accurate final results despite using
relatively short expansions~(\ref{Psilc}).

To control the numerical uncertainty of the shielding constant, we performed calculations 
with several wave functions with regularly increased expansion, i.e., $N=256,\; 384,\; 512,\; 768...$. 
At the stage of the optimization, the intermediate states were assumed to be of the same size as
the wave function $\Psi$. 
For each $N$, two separate optimizations were performed -- the goal function was of the same form
as the second term of the isotropic shielding constant $\sigma$ of Eq.~(\ref{isosh}), but
the second-order expression was made symmetric, i.e., both Hamiltonians were either $H_B^i$ or $H_I^i$
[see Eqs.~(\ref{HB}) and~(\ref{HI}) for their definitions].
In the final calculations, the two optimized basis sets were added together, forming an intermediate
state function of size $2\,N$.

Results of the shielding constant calculations for HD and HT, performed using the DNA method, 
are presented in Table~\ref{DNAresultsHD}.
The uncertainties of the extrapolated values reflect the numerical convergence only and do not
account for the missing relativistic effects of the relative order $\alpha^2$. The numerical
accuracy of $\sigma$ is estimated as better than $6\cdot 10^{-8}$, whereas that of $\delta\sigma$ as $8\cdot10^{-5}$.

\subsection{Numerical calculations in the NAPT framework}

First, all of the matrix elements in the shielding difference $\delta\sigma(R)$ 
in Eq.~\eqref{dsigmaR} are converted to the forms without the $R$-derivatives
acting on the electronic wave function. Taking $\nabla_R^i[Q^j]=\nabla_R^i[Q'^j]=0$, one obtains
\begin{align}\nonumber
&\left\langle \vec Q'\,\frac{1}{{\cal E}_{\rm el}-H_{\rm el}}\,\vec Q\times(\vec\nabla_R-\cev\nabla_R)\right\rangle_{\rm el}\\ 
&= \epsilon^{ijk}\biggl[
-\left\langle\nabla_R^k[V-{\cal E}_{\rm el}]\,\frac{1}{({\cal E}_{\rm el}-H_{\rm el})'}\,Q'^i\,\frac{1}{{\cal E}-H}\,Q^j\right\rangle
\nonumber \\
&\hspace{7.3ex}-\left\langle Q'^i\,\frac{1}{{\cal E}_{\rm el}-H_{\rm el}}\,\nabla^k_R[V-{\cal E}_{\rm el}]\,\frac{1}{{\cal E}_{\rm el}-H_{\rm el}}\,Q^j\right\rangle \nonumber \\
&\hspace{7.3ex}+\left\langle Q'^i\,\frac{1}{{\cal E}_{\rm el}-H_{\rm el}}\,Q^j\,\frac{1}{({\cal E}_{\rm el}-H_{\rm el})'}\, \nabla^k_R[V-{\cal E}_{\rm el}]\right\rangle\biggr],
\end{align}
%
where all matrix elements are expressed in terms of electronic operators.
This conversion allows additional numerical evaluation of the radial derivatives to be circumvented, 
which is problematic in accurate numerical calculations. 

The ground electronic state wave function is represented as a linear combination of two-center 
ECG basis functions expressed in terms of interparticle coordinates,
\begin{equation}
  \varphi_{\Sigma^+} = e^{-a_{1A}\,r_{1A}^2 -a_{1B}\,r_{1B}^2 -a_{2A}\,r_{2A}^2 -a_{2B}\,r_{2B}^2 - a_{12}\,r_{12}^2 },
\end{equation}
where indices $A,B$ and $1,2$ are related to nuclei and electrons. 
Basis functions $\varphi$ are properly symmetrized to represent the singlet gerade electronic state,
\begin{eqnarray}\label{lexp}
\phi &=& \sum_i^N c_i  (1  + P_{A\leftrightarrow B})\,(1+P_{1\leftrightarrow 2})\,\varphi_{\Sigma^+,i}\,,
\end{eqnarray}
where $P_{i\leftrightarrow j}$ is the particle exchange operator and $c_i$ is a linear variational parameter.
Additional basis functions are necessary for calculations of the matrix elements containing $1/({\cal E}-H)$
resolvents with the intermediate states of $\Sigma^-$ and $\Pi$ symmetry. The following functions
were employed for this purpose
\begin{eqnarray} 
  \varphi_{\Sigma^-} &=& \vec R \cdot (\vec r_{1A} \times \vec r_{2A})\,\varphi_{\Sigma^+}\,, \\ 
 \vec\varphi_{\Pi} &=& (\vec R \times \vec r_{1A})\,\varphi_{\Sigma^+}.
\end{eqnarray}
Variational calculations are performed using $N=128$ and $256$ expansions~(\ref{lexp}). 
For the given $N$, the optimization was also performed on the intermediate  $\Pi$ and $\Sigma^{-}$ states using symmetric second-order expressions with the same basis size $N$. For ${\Sigma^+}$ intermediate states,
we use a fixed sector of basis functions with non-linear parameters taken from the $\Sigma^+$ wave function
of size $N/2$. Such a combination improves the quality of the electronic ground state, which must be
precisely removed from the reduced resolvent $1/({\cal E}-H)'$. Numerical values of the matrix elements were
checked against the $R\rightarrow\infty$  limit. The known separated-atoms 
limit for the shielding constant is
\begin{eqnarray}
\sigma_A^{(1)}(\infty) &=& -\frac{\alpha^2}{3}\,\frac{m}{m_A}\,\biggl(1+\frac{g_A-1}{g_A}\biggr)\,, \\[1.2ex]
\delta\sigma_{AB}(\infty) &=& \sigma_B^{(1)}(\infty) - \sigma_A^{(1)}(\infty)\,,
\end{eqnarray}
\begin{figure}[!hb]
\centerline{\includegraphics[scale=0.42]{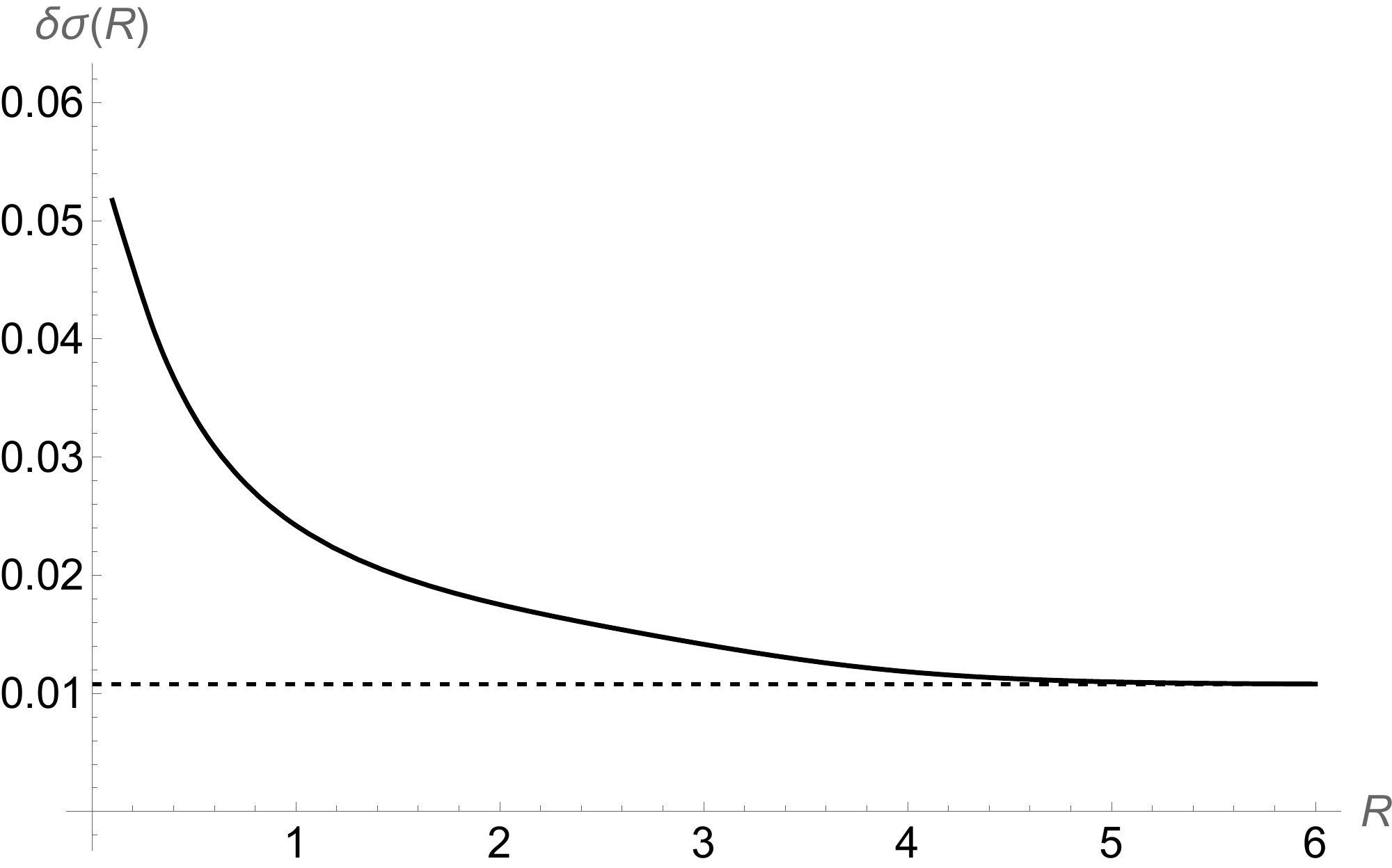}}
\caption{\label{figdsigmaHD}
The difference $\delta\sigma({\rm HD},R)$ in ppm 
         of the shielding constant between the deuteron and the proton in HD 
         as a function of the internuclear distance $R$ in a.u. 
         The horizontal line is the separated-atoms limit.
         }
\end{figure}
\begin{figure}[!ht]
\centerline{\includegraphics[scale=0.42]{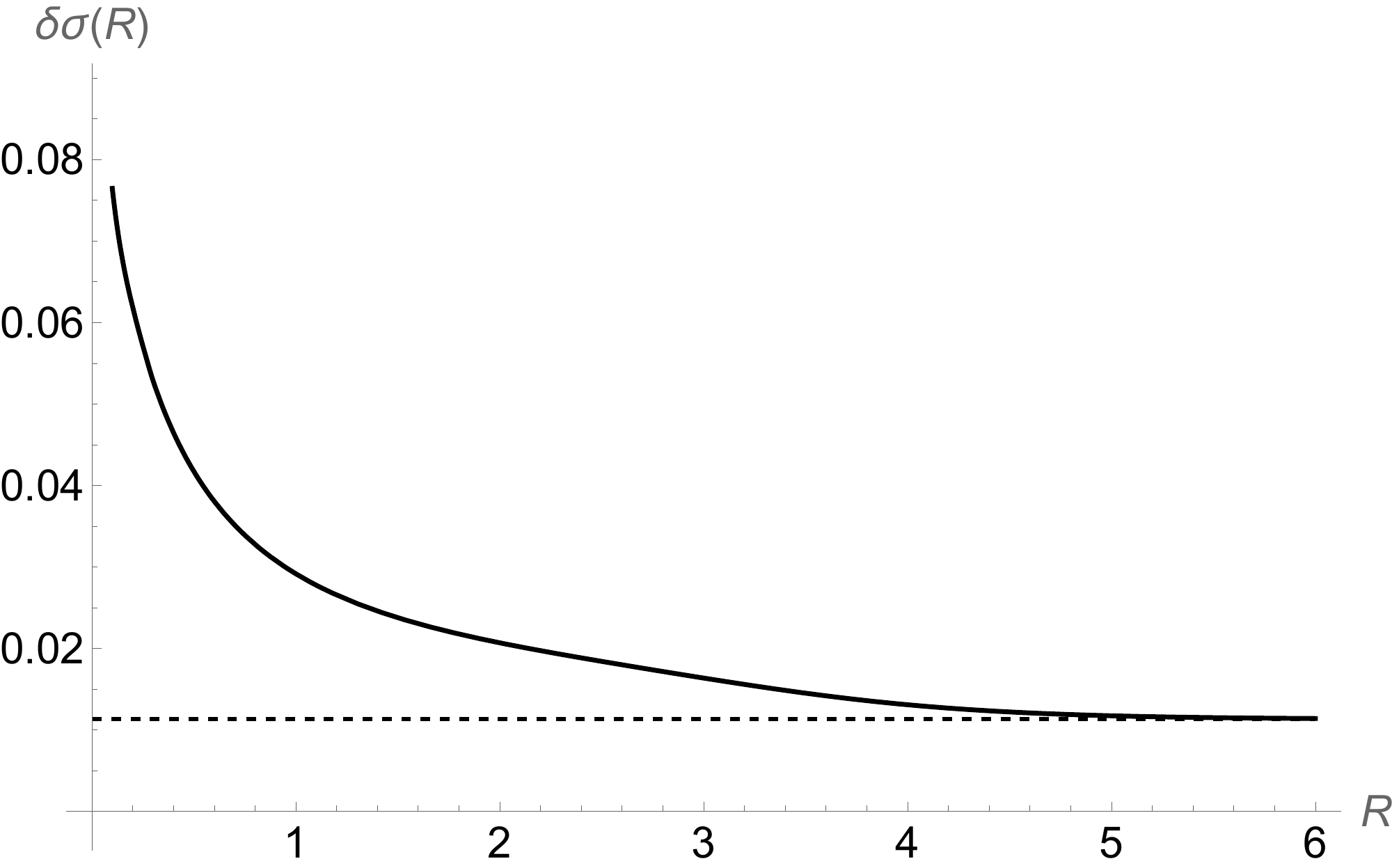}}
\caption{ \label{figdsigmaHT}
The difference $\delta\sigma({\rm HT},R)$ in ppm 
         of the shielding constant between the triton and the proton in HT 
         as a function of the internuclear distance $R$ in a.u. 
         The horizontal line is the separated-atoms limit.
         }
\end{figure}
which yields the following numerical values: $\delta\sigma_{HD}(\infty)\simeq0.010\,753\cdot10^{-6}$ 
and $\delta\sigma_{HT}(\infty) \simeq 0.011\,326 \cdot 10^{-6}$. 
Numerical results in the range $R\in\langle 0.1,6\rangle$ a.u. are collected in Appendix~\ref{app:table}
and presented graphically in Figs \ref{figdsigmaHD} and \ref{figdsigmaHT}, where the correct
behavior at the $R\to\infty$ limit can be noted. The final results for the nuclear magnetic shielding 
constants are obtained by averaging with the nuclear function according to Eq.~(\ref{eq:sigmachi}).
Apart from checking the consistency with the DNA calculations in the ground energy level, 
we also obtain the shielding constant differences for excited rotational states, which are populated 
in 300 K temperature in which the corresponding experiments were performed.

\section{Results and conclusions}

\begin{table}[!ht]
 \caption{\label{Tresults} 
 Magnetic moment of the deuteron $\mu_d$ and the triton $\mu_t$ determined from the proton
 magnetic moment $\mu_p$ using Eq.~(\ref{eq:muAmuB}). $\mu_x({\rm HX})$  denotes 
 the shielded magnetic moment of the nucleus $x$ in the HX molecule. $\Delta_T\delta\sigma$
 is the change in $\delta\sigma$ caused by the temperature averaging. 
 NAPT results have explicit numerical uncertainties and implicit ones of relative order $3\cdot 10^{-2}$ due to nonadiabatic effects.
 The uncertainty of the final shielding $\delta\sigma({\rm HX},300K)$ due to relativistic effects is estimated to be $1\cdot 10^{-6}$.}
 \begin{ruledtabular}
 \begin{tabular}{l x{1.18} l}
  Quantity                        & \centt{Value}\qquad            & Reference  \\ 
\hline
%
$\mu_p$                                   & 2.792\,847\,344\,62(82)\,\mu_N &\cite{Schneider:17} \\[2ex]
$\delta\sigma({\rm HD},v=0,J=0)$          & 0.020\,433\cdot 10^{-6} & NAPT \\
$\delta\sigma({\rm HD},v=0,J=0)$          & 0.019\,901(1)\cdot 10^{-6} & DNA \\
$\Delta_T\delta\sigma({\rm HD},300K)$     &-0.000\,023\,7\cdot 10^{-6} & NAPT \\
$\delta\sigma({\rm HD},300K)$             & 0.019\,877(1)\cdot 10^{-6} & DNA+$\Delta_T$ \\
                                          & 0.020\,20(2)\cdot 10^{-6} & \cite{Puchalski:15} \\[2ex]
%
$\mu_p({\rm HD})/\mu_d({\rm HD})$         & 3.257\,199\,516(10)        &\cite{Neronov:12,CODATA:18} \\
$\mu_d = \mu_d({\rm HD})/\mu_p({\rm HD})$ & 0.857\,438\,233\,8(26)\,\mu_N & This work\\
$\qquad\times(1+\delta\sigma)\,\mu_p$     & 0.857\,438\,233\,8(22)\,\mu_N & CODATA \cite{CODATA:18}\\ 
                                          & 0.857\,438\,234\,6(53)\,\mu_N & \cite{Puchalski:15} \\[2ex]
%
$\delta\sigma({\rm HT},v=0,J=0)$          & 0.024\,368\cdot 10^{-6} & NAPT \\
$\delta\sigma({\rm HT},v=0,J=0)$          & 0.023\,975(2)\cdot 10^{-6} & DNA \\
$\Delta_T\delta\sigma({\rm HT},300K)$     &-0.000\,030\,0\cdot 10^{-6} & NAPT \\
$\delta\sigma({\rm HT},300K)$             & 0.023\,945(2)\cdot 10^{-6} & DNA+$\Delta_T$ \\
                                          & 0.024\,14(2)\cdot 10^{-6} & \cite{Puchalski:15} \\[2ex]
$\mu_t({\rm HT})/\mu_p({\rm HT})$         & 1.066\,639\,893\,3(21) &\cite{Neronov:11,CODATA:18}\\ 
$\mu_t = \mu_t({\rm HT})/\mu_p({\rm HT})$ & 2.978\,962\,465\,0(59)\,\mu_N & This work \\
$\qquad\times(1+\delta\sigma)\,\mu_p$     & 2.978\,962\,465\,6(59)\,\mu_N & CODATA \cite{CODATA:18} \\ 
                                          & 2.978\,962\,471(10)\,\mu_N & \cite{Puchalski:15} 
 \end{tabular}
 \end{ruledtabular}
\end{table}

Our final numerical results of the deuteron and triton magnetic moments are presented 
in Table II. As described above, the shielding difference $\delta\sigma= \sigma_{d/t} - \sigma_p$ 
for the ground rovibrational level ($v=0,J=0$) was obtained in two manners;
using NAPT, which includes only the leading nonadiabatic effects, and using the DNA method,
which completely accounts for the finite nuclear mass effects. The knowledge of the latter
enables estimation of the missing higher-order finite-mass effects in the former. 
The difference NAPT--DNA corresponds to the relative uncertainty of $3\cdot10^{-2}$ and is 
consistent with estimation by the square root of the inverse power of the nuclear masses.
This difference, however, is an order of magnitude larger than our previous estimation 
by $\me/\mun\sim10^{-3}$ in Ref. \cite{Puchalski:15}. 

The DNA value is augmented by the temperature averaging  correction $\Delta_T\delta\sigma$
evaluated for $T=300$ kelvins -- the temperature in which the measurements were performed. 
Since it is a quite small effect, for its calculation the adiabatic rotational energies $\varepsilon_J$ and wave
functions $\chi_J$, obtained in the NAPT framework, were employed.
First, the shielding difference $\delta\sigma(R)$ was averaged for the lowest 
10 rotational states, 
$\langle\delta\sigma\rangle_{J}=\langle\chi_J|\delta\sigma|\chi_J\rangle$
and the rotational energies were used to obtain the Boltzmann weights
\begin{align}
w_J&=\dfrac{(2J+1)\exp{[-\varepsilon_J/(kT)]}}{\sum_J(2J+1)\exp{[-\varepsilon_J/(kT)]}}.
\end{align}
Then the rotationally averaged $\delta\sigma(R)$ values were summed up
with the $J=0$ value subtracted,
\begin{align}
\Delta_T\delta\sigma&=\sum_J w_J\,\langle\delta\sigma\rangle_{J}
 -\langle\delta\sigma\rangle_{0}\,.
\end{align}
The relative uncertainty of $\Delta_T\delta\sigma$ comes mainly from nonadiabatic effects and is about $3\cdot 10^{-2}$,
while numerical uncertainty is completely negligible.

The new shielding values slightly differ from our previous result
due to underestimation of the nonadiabatic effects and due to mistakes in the final formulas, 
which we have already discussed. 
The final shielding difference $\delta\sigma$, labeled "DNA+$\Delta_T$" in Table~\ref{Tresults}, 
was used in the evaluation of the magnetic moment of the nucleus $x$ according to 
\begin{equation}
\mu_x = \mu_x({\rm HX})/\mu_p({\rm HX}) (1+\delta\sigma)\,\mu_p
\end{equation}
derived directly from Eq.~(\ref{eq:muAmuB}).

The obtained values of the deuteron and the triton magnetic moments 
do not differ greatly from the the CODATA 2018 values which used our previous results.
Their accuracy is limited exclusively by experimental uncertainties in the NMR determination 
of the magnetic moment ratio. In principle, the deuteron and triton magnetic moments 
can be obtained as accurately as that of the proton, provided the experimental uncertainty 
in the NMR frequency ratio is reduced by a factor of 10.

Similarly, the magnetic moments of all stable nuclei can, in principle, be determined by a chain 
of NMR measurements through $^3$He. The $^3$He magnetic moment can be obtained by 
measuring the magnetic moment ratio in H$_2$ and $^3$He. For this, one would need nonadiabatic shielding 
and relativistic correction in H$_2$, while for $^3$He the shielding is already accurately known 
\cite{Wehrli:21}. The nonadiabatic shielding in H$_2$ and other molecules can be calculated 
by means of the DNA method or NAPT, as presented in this work, while relativistic corrections 
are yet to be calculated in a similar way as for $^3$He. Such results would eventually allow 
for much-improved determination of all the nuclear magnetic moments.

\begin{acknowledgments}
This work has been supported by National Science Center (Poland) Grants No. 2016/23/B/ST4/01821 and No. 2019/34/E/ST4/00451, 
as well as by a computing grant from the Pozna\'n Supercomputing and Networking Center and by PL-Grid Infrastructure.
A.S. acknowledges additional support by Grant no. POWR.03.02.00-00-I020/17 co-financed by the European Union through the European Social Fund under the Operational Program Knowledge Education Development.
\end{acknowledgments}

\appendix


\section{Operator matrix element in NAPT}
\label{app:NAPT}
Although most of the consideration below will be valid for an arbitrary molecule, to be more specific,
we consider a two-electron diatomic molecule. The total wave function $\psi$ is a solution of the stationary \SE
\begin{equation}
H_0\,\Psi = E\,\Psi\,, \label{ESE}
\end{equation}
with the Hamiltonian
\begin{equation}
H_0 = H_{\rm el} + H_{\rm n}\,, \label{EH}
\end{equation}
split into the electronic and nuclear parts.  In the electronic Hamiltonian
\begin{equation}
H_{\rm el} = -\suma\frac{\nabla^2_a}{2\,\me} + V\,, \label{EHel}
\end{equation}
where $V$ is the Coulomb interaction potential
\begin{equation}
V = -\frac{1}{r_{1A}}-\frac{1}{r_{1B}}-\frac{1}{r_{2A}}-\frac{1}{r_{2B}}
+\frac{1}{r_{12}}+\frac{1}{R},
\end{equation}
the nuclei have fixed positions $\vec R_A$ (proton) and $\vec R_B$ (deuteron/triton), 
and $\vec R = \vec R_A-\vec R_B$. The nuclear Hamiltonian is
\begin{align}
H_{\rm n} =&\ -\frac{\nabla_A^2}{2\,m_A} -  \frac{\nabla_B^2}{2\,m_B}\,.\label{EHn}
\end{align}
Because $m/m_A$ and $m/m_B$ are small, it is customary to assume that the total wave function of the molecule
\begin{equation}
\psi(\vec r,\vec R) = \phi(\vec r;\,\vec R) \; \chi(\vec R) 
\end{equation}
is a product of the electronic wave function $\phi$ that depends parametrically on $R$,
and the nuclear wave function $\chi$. 
The electronic wave function obeys the clamped nuclei electronic Schr\"odinger equation
\begin{equation}
\bigl[H_{\rm el}-\Eel(R)\bigr]\,|\phi\rangle = 0, \label{ESEel}
\end{equation}
while the wave function $\chi$ is a solution to the nuclear Schr\"odinger 
equation with the effective potential generated by electrons
\begin{equation}
({\cal H}-E)\,|\chi\rangle = 0\,, \label{ESEn}
\end{equation}
where 
\begin{align}\label{Ead}
{\cal H} &= -\frac{\vec\nabla_R^2}{2\,\mun}+\Ea(R)+\Eel(R). 
\end{align}
The function 
\begin{equation}
\Ea(R)=\EVS{H_{\rm n}}
\end{equation} 
is the so-called adiabatic correction, where the subscript "el" is explained in the following.
We shall consider two different types of matrix elements of an operator $\Omega$ containing differentiation over $R$,
which differ in the range of differentiation. The first type of the matrix element
\begin{align}
\langle\phi'|\Omega|\phi\rangle
\end{align} 
will be understood as an operator acting in the subspace of rotational and vibrational states $\chi$\begin{align}
\langle\chi'|\langle\phi'|\Omega|\phi\rangle|\chi\rangle\,,
\end{align} 
which means that $\Omega$ acts on both $\phi$ and $\chi$. The second type of the matrix element
\begin{align}
\langle\phi'|\Omega|\phi\rangle_\mathrm{el}\,,
\end{align}
distinguished by the subscript "el", has the differentiation range limited to the single function
$\phi$ immediately following $\Omega$.
For example,
\begin{align}
\langle\phi'|\nr|\phi\rangle_\mathrm{el} = \langle\phi'|\nr\phi\rangle\,.
\end{align}
To shorten the forthcoming expressions, we define the "left-hand" differential operator,
\begin{align}
\langle\phi'|\rn|\phi\rangle_\mathrm{el} = \langle\nr\phi'|\phi\rangle.
\end{align}
For a clear definition of the scope of action of the $R$ derivative, we introduce another
symbol
\begin{equation}\label{EnrQ}
\nr[Q]\equiv[\nr,Q]\,.
\end{equation}
For example, for arbitrary states $\phi'$, $\phi$, we have
\begin{align}
\nr[\langle\phi'|\phi\rangle] &=\langle\phi'|\rn + \nr | \phi\rangle_\mathrm{el} \,.
\end{align}
If these states are orthogonal $\langle\phi'|\phi\rangle = 0$, then
\begin{align}
\langle\phi'|\rn + \nr | \phi\rangle_\mathrm{el} =0 \label{Elrdiff}
\end{align}
and the left- and right-hand derivatives differ by sign.
If, in turn, $\phi' = \phi$ and $\phi$ is a normalized real function $(\langle\phi|\phi\rangle = 1)$, 
then these derivatives vanish,
\begin{align}
\langle\phi|\cev{\nabla}_R  | \phi\rangle_\mathrm{el}  = 
\langle\phi|\vec{\nabla}_R  | \phi\rangle_\mathrm{el} = 0\,.
\end{align}
Therefore, for these states the matrix element of the nuclear kinetic energy is 
\begin{align}
\langle\phi|\vec\nabla_R^2|\phi\rangle =&\ \vec\nabla_R^2 + \langle\phi|\vec\nabla_R^2|\phi\rangle_\mathrm{el}
\nonumber \\ =&\  
\vec\nabla_R^2 - \langle\phi | \cev\nabla_R\,\vec\nabla_R | \phi\rangle_\mathrm{el}\,,
\end{align} 
which explains the form of the nuclear Hamiltonian in Eq. (\ref{Ead}).

We will be using the following type of matrix element with arbitrary real functions $\phi$ and $\phi'$:
\begin{align}
\vec Z \equiv \langle\phi'|\vec\nabla_R|\phi\rangle  + \langle\phi|\cev\nabla_R|\phi'\rangle.
\end{align}
This expression can be transformed as follows:
\begin{align}
\vec Z =&\  \langle\phi'|\vec\nabla_R|\phi\rangle_{\rm el}  + \langle\phi|\cev\nabla_R|\phi'\rangle_{\rm el}
+\langle\phi'|\phi\rangle\,\vec\nabla_R  + \cev\nabla_R\,\langle\phi|\phi'\rangle
\nonumber \\ =&\ 
2\,\langle\phi'|\vec\nabla_R|\phi\rangle_{\rm el} - [\vec\nabla_R\,,\, \langle\phi'|\phi\rangle]
\nonumber \\ =&\  \langle\phi'|\vec\nabla_R-\cev\nabla_R|\phi\rangle_{\rm el}
\end{align}
If, additionally, $\langle\phi'|\phi\rangle=0$, then $\vec Z = 2\,\langle\phi'|\nr|\phi\rangle_{\rm el} $.

In the reference frame attached to the geometrical center of the nuclei, $H_{\rm n}$
can be written as a sum of two components,
\begin{align}
H'_{\rm n} =&\ - \frac{\nabla^2_{\!R}}{2\,\mun}
           - \frac{{\nabla}_{\!\mathrm{el}}^2}{8\,\mun}\\
H''_{\rm n} =&\ \frac{1}{2}\biggl(\frac{1}{M_A}-\frac{1}{M_B}\biggr)\,
             \vec\nabla_R\cdot\nel
\end{align}
with the first one being even and the second- one being odd with respect to the inversion.
In the above $\nel=\suma\na$ and $\mun=\left(1/M_A+1/M_B\right)^{-1}$ 
is the reduced nuclear mass.
Due to the inversion symmetry of $\phi$ with respect to the geometrical center,
$\langle\phi|H''_{\rm n}|\phi\rangle_{\rm el}=0$ and
\begin{align}
\Ea(R)&=\EVS{H'_{\rm n}} \nonumber\\
&=\frac{1}{2\,\mun}\,\left\langle\phi\left| \cev\nabla_R\cdot\vec\nabla_R\right|\phi\right\rangle_{\rm el}
-\frac{1}{8\,\mun}\left\langle\phi\left|\nel^2\right|\phi\right\rangle.
\end{align}

However, for the determination of the difference in the shielding of the proton and deuteron magnetic moments, 
we shall consider the reference frame centered on one of the nuclei, say nucleus $A$.
The nuclear Hamiltonian then becomes
\begin{align}
H_{\rm n} =&\ -\frac{(\vec\nabla_B + \vec\nabla_{\rm el})^2}{2\,m_A} -\frac{\vec\nabla_B^2}{2\,m_B} \label{hnall}
\nonumber \\ =&\ H'_{\rm n} + H''_{\rm n}
\end{align}
where
\begin{align}
H'_{\rm n} =&\  -\frac{\vec\nabla_B^2}{2\,\mun}\\
H''_{\rm n} =&\ -\frac{\vec\nabla_{\rm el}^2}{2\,m_A} -\frac{\nel\cdot\nb}{m_A}\,.\label{eq:Hnbis}
\end{align}
Vectors with the origin at $A$, pointing at a particle $a$ will be denoted by $\vec{x}_a$, and $\nb\equiv-\nr$.
The diagonal matrix element of  $H'_{\rm n}$ 
\begin{align}
\langle\phi| H'_{\rm n} |\phi\rangle_{\rm el} =  \frac{1}{2\,\mun}\,
\langle\phi|\bn\cdot\nb|\phi\rangle_{\rm el} \,. 
\end{align}
This term does not contribute to the difference in the nuclear magnetic shielding
because it depends on the reduced nuclear mass only.
The diagonal matrix element of the second term in $H''_{\rm n}$ is
\begin{align}
\delta \cal{E}&=-\frac{1}{m_A}\,\langle\phi|\nel\cdot\nb|\phi\rangle_\mathrm{el}\,.
\end{align}
By acting $\vec\nabla_B$ on the Schr\"odinger equation (\ref{ESEel}),
$\delta \cal{E}$  can be transformed to
\begin{align}\label{EdE1}
\delta \cal{E}&=-\frac{1}{m_A}\,\langle\phi|\nel\frac{1}{({\cal E}_{\rm el}-H_{\rm el})'}\nb[V-{\cal E}_{\rm el}]|\phi\rangle\,,
\end{align}
where the notation introduced in Eq.~(\ref{EnrQ}) was applied.
To make formulas more compact, we now introduce an abbreviated notation 
$\langle\dots\rangle\equiv\langle\phi|\dots|\phi\rangle$.
Because $\nel = \me\, [\vec x_{\rm el}\,,\,H_{\rm el}]$, where $\vec{x}_{\rm el}=\sum_a\vec{x}_a$, 
the expectation value~(\ref{EdE1}) can be rewritten as
\begin{align}
\delta {\cal E}&=
\frac{\me}{m_A}\,\langle[\vec x_{\rm el}\,,\,{\cal E}_{\rm el} - H_{\rm el}]\frac{1}{({\cal E}_{\rm el}-H_{\rm el})'}\nb[V-{\cal E}_{\rm el}]\rangle\nonumber \\
&=\frac{\me}{m_A}\,\langle\vec x_{\rm el}\,(I-|\phi\rangle\,\langle\phi|)\,\nb[V-{\cal E}_{\rm el}]\rangle\nonumber \\
&=\frac{\me}{m_A}\,\langle\left(\vec x_{\rm el}-\langle\vec x_\mathrm{el}\rangle\right)\,\nb[V-{\cal E}_{\rm el}]\rangle
\end{align}
and we note that $\langle\nb[V-{\cal E}_{\rm el}]\rangle = 0$.
It is convenient to define the following operator
\begin{align}
\tilde H_{\rm n} \equiv
(\vec x_{\rm el}-\langle\vec x_\mathrm{el}\rangle)\,\nb[V-{\cal E}_{\rm el}]-\frac{\vec\nabla_{\rm el}^2}{2\,\me}\,,
\end{align}
which will be used in the calculation of the shielding constant difference. 
Because the adiabatic energy does not depend on the reference frame, the diagonal matrix element 
of $\tilde H_{\rm n}$, with $\langle\vec x_\mathrm{el}\rangle = -\vec R$, vanishes:
\begin{align}
\langle\phi|H''_{\rm n}|\phi\rangle_{\rm el}=\frac{\me}{m_A}\,\langle\tilde{H}_{\rm n}\rangle=0\,,
\end{align}
which can be shown by replacing $\vec\nabla_B
= (\vec\nabla_A+\vec\nabla_B)/2 +(\vec\nabla_B-\vec\nabla_A)/2$.
The expectation value of $\tilde H_{\rm n}$ with the second term vanishes
because the ground state has gerade symmetry, while the first term can be
replaced by $-\vec\nabla_{\rm el}/2$; hence,
\begin{align}
\langle\tilde H_{\rm n}\rangle
&=\left\langle -\frac{1}{2}(\vec x_{\rm el}-\langle\vec x_\mathrm{el}\rangle)\,[\vec\nabla_{\rm el}\,,\,H_{\rm el}-{\cal E}_{\rm el}]-\frac{\vec\nabla_{\rm el}^2}{2\,\me}\right\rangle \nonumber \\
&=\left\langle \frac{1}{2}[\vec x_{\rm el}-\langle\vec x_\mathrm{el}\rangle\,,\,
H_{\rm el}-{\cal E}_{\rm el}]\cdot\,\vec\nabla_{\rm el}-\frac{\vec\nabla_{\rm el}^2}{2\,\me}\right\rangle = 0\,.
\end{align}

\section{Derivation of finite nuclear mass corrections $\sigma_{\rm n}$}
\label{app:sigman}
The shielding correction $\sigma_{\rm n}$ is obtained from the leading one by correcting 
all matrix elements by $H''_{\rm n}$ in Eq. (\ref{eq:Hnbis}), and
it is split into two terms, i.e., $\sigma_{\rm n} = \sigma_\mathrm{n1}+\sigma_\mathrm{n2}$.
Consider the finite nuclear mass corrections to the first BO term
\begin{align}
\sigma_\mathrm{n1} =&\ 
\frac{\alpha^2}{3}\,\biggl[
\Big\langle Q_0\,\frac{1}{({\cal E}_{\rm el}-H_{\rm el})'}\,{H''_{\rm n}}\Big\rangle
+ \Big\langle {H''_{\rm n}}\,\frac{1}{({\cal E}_{\rm el}-H_{\rm el})'}\,Q_0 \Big\rangle\biggr],
\end{align}
where
\begin{align}
Q_0 = \sum_{b}\frac{1}{x_b}.
\end{align}
$\vec\nabla_{\rm el}^2$, the first term in  $H''_{\rm n}$ of Eq.~(\ref{eq:Hnbis}), does not need 
any further transformation, while the second term does:
\begin{align}
{\delta\sigma_\mathrm{n1}} =&\ 
-\frac{\alpha^2}{3\,m_A}\,\biggl[
\Big\langle Q_0\,\frac{1}{({\cal E}_{\rm el}-H_{\rm el})'}\,\vec\nabla_B\vec\nabla_{\rm el} \Big\rangle
\nonumber \\&\
+ \Big\langle \cev\nabla_B\cev\nabla_{\rm el}\,\frac{1}{({\cal E}_{\rm el}-H_{\rm el})'}\,Q_0\Big\rangle\biggr].
\end{align} 
This is the $Z$ type matrix element, so
\begin{widetext}
\begin{align}
{\delta\sigma_\mathrm{n1}} =&\ 
-\frac{\alpha^2}{3\,m_A}\,
\Big\langle Q_0\,\frac{1}{({\cal E}_{\rm el}-H_{\rm el})'}\,(\vec\nabla_B-\cev\nabla_B)\vec\nabla_{\rm el} \Big\rangle_{\rm el}
\nonumber \\ =&\ 
-\frac{\alpha^2\,m}{3\,m_A}
\Big\langle Q_0\,\frac{1}{({\cal E}_{\rm el}-H_{\rm el})'}\,[{\cal E}_{\rm el}-H_{\rm el}\,,\,\vec x_{\rm el} - \langle \vec x_{\rm el}\rangle]\,(\vec\nabla_B-\cev\nabla_B) \Big\rangle_{\rm el}
\nonumber \\ =&\ 
-\frac{\alpha^2\,m}{3\,m_A}
\Big\langle\bigl(Q_0-\bigl\langle Q_0 \bigr\rangle\bigr)\,(\vec x_{\rm el} - \langle \vec x_{\rm el}\rangle)\,(\vec\nabla_B-\cev\nabla_B) \Big\rangle_{\rm el}
+ \frac{\alpha^2\,m}{3\,m_A}
\Big\langle Q_0\,\frac{1}{({\cal E}_{\rm el}-H_{\rm el})'}\,(\vec x_{\rm el} - \langle \vec x_{\rm el})\, [{\cal E}_{\rm el}-H_{\rm el}\,,\,\vec\nabla_B-\cev\nabla_B] \Big\rangle_{\rm el}
\nonumber \\ =&\ 
\frac{2\,\alpha^2\,m}{3\,m_A}
\Big\langle Q_0\,\frac{1}{({\cal E}_{\rm el}-H_{\rm el})'}\,(\vec x_{\rm el} - \langle \vec x_{\rm el})\, \vec\nabla_B[V-\cal{E}_{\rm el}] \Big\rangle\,.
\end{align}
\end{widetext}
Therefore,
\begin{align}
\sigma_\mathrm{n1} =&\ 
\frac{2\,\alpha^2}{3\,m_A}
\langle\phi| Q_0\,\frac{1}{({\cal E}_{\rm el}-H_{\rm el})'}\, {\tilde H}_{\rm n} |\phi\rangle\,.
\end{align}
Consider now the finite nuclear mass corrections to the second BO term, and let 
\begin{align}
\vec Q_1 =&\  \sum_a\vec x_a\times\vec p_a,\qquad\langle\vec Q_1\rangle = 0,\\
\vec Q_2 =&\ \sum_{b}\frac{\vec x_b\times\vec p_b}{x_b^3},\qquad\langle\vec Q_2\rangle = 0\,.
\end{align}
Then
\begin{widetext}
\begin{align}
\sigma_\mathrm{n2} =&\ \frac{\alpha^2}{3}\,\biggl[
\Big\langle {H''_{\rm n}}\,\frac{1}{({\cal E}_{\rm el}-H_{\rm el})'}\,
\vec Q_1 \frac{1}{({\cal E}_{\rm el}-H_{\rm el})'} \vec Q_2\Big\rangle
+\Big\langle \vec Q_1\frac{1}{({\cal E}_{\rm el}-H_{\rm el})'}\, H''_{\rm n}\frac{1}{({\cal E}_{\rm el}-H_{\rm el})'}\,\vec Q_2\Big\rangle
\nonumber \\ &\ 
+\Big\langle \vec Q_1\,\frac{1}{({\cal E}_{\rm el}-H_{\rm el})'}\,\vec Q_2\,
\frac{1}{({\cal E}_{\rm el}-H_{\rm el})'} \,{H''_{\rm n}}\Big\rangle\biggr]. \label{shield_ad}
\end{align}
One notes that the matrix element due to the second term in $H''_{\rm n}$ is of  the $Z$-type, so
\begin{align}
{\delta\sigma_\mathrm{n2}} =&\ -\frac{\alpha^2}{6\,m_A}\,\biggl[
\Big\langle (\vec\nabla_B-\cev\nabla_B)\,\vec\nabla_{\rm el}\,\frac{1}{({\cal E}_{\rm el}-H_{\rm el})'}\,\vec Q_1
\frac{1}{({\cal E}_{\rm el}-H_{\rm el})'} \vec Q_2 \Big\rangle_{\rm el}
+\Big\langle \vec Q_1\frac{1}{({\cal E}_{\rm el}-H_{\rm el})'}\, (\vec\nabla_B-\cev\nabla_B)\,\vec\nabla_{\rm el}\,
\frac{1}{({\cal E}_{\rm el}-H_{\rm el})'}\,\vec Q_2\Big\rangle_{\rm el}
\nonumber \\ &\ 
+\Big\langle \vec Q_1\,\frac{1}{({\cal E}_{\rm el}-H_{\rm el})'}\, \vec Q_2\,
\frac{1}{({\cal E}_{\rm el}-H_{\rm el})'} \,(\vec\nabla_B-\cev\nabla_B)\,\vec\nabla_{\rm el}\Big\rangle_{\rm el}\biggr].
\end{align}
Using commutation relation similar to those for $\delta_1\sigma$, one obtains
\begin{align}
\sigma_\mathrm{n2} &=\frac{\alpha^2}{3}\,\frac{m}{m_A}\,\biggl[
\Big\langle {\tilde H}_{\rm n}\,\frac{1}{({\cal E}_{\rm el}-H_{\rm el})'}\,
\vec Q_1\frac{1}{({\cal E}_{\rm el}-H_{\rm el})'}\vec Q_2\Big\rangle
+\Big\langle \vec Q_1
\frac{1}{({\cal E}_{\rm el}-H_{\rm el})'}\, {\tilde H}_{\rm n}
\frac{1}{({\cal E}_{\rm el}-H_{\rm el})'}
\vec Q_2\Big\rangle
\nonumber \\ &\ 
+\Big\langle \vec Q_1\,\frac{1}{({\cal E}_{\rm el}-H_{\rm el})'}\,\vec Q_2\frac{1}{({\cal E}_{\rm el}-H_{\rm el})'} {\tilde H}_{\rm n}\Big\rangle
-\biggl\langle\vec x_\mathrm{el}\times\vec p_B\,\frac{1}{({\cal E}_\mathrm{el}-H_\mathrm{el})'}\vec Q_2\biggr\rangle_{\rm el}
-\biggl\langle \vec Q_1\,\frac{1}{({\cal E}_\mathrm{el}-H_\mathrm{el})'}\sum_b \frac{\vec x_b}{x_b^3}\times \vec p_B\biggr\rangle_{\rm el}
\biggr].
\end{align}
where $\vec p_B$ is in the symmetrized form given by Eq. (\ref{29}).
This concludes the derivation of the $\sigma_\mathrm{n}$ term of Eq.~(\ref{eq:sigman}).
\end{widetext}

\section{Numerical values of $\delta\sigma(R)$ functions}
\label{app:table}

Table~\ref{NAPTresultsHD} contains numerical values of the difference in the nuclear magnetic shielding $\delta\sigma(R)$
in HD and in HT, evaluated using a 256-term ECG wave function. Cubic spline interpolation
and an inverse power expression were used to perform interpolation at short
and long range of the internuclear distance $R$, respectively. Interpolated functions 
were subsequently employed in vibrational and thermal averaging.
\begin{table}[!hb]
 \caption{$\delta\sigma(\mathrm{HX})=\sigma_x(\mathrm{HX})-\sigma_p(\mathrm{HX})$ 
 (in ppm) evaluated using NAPT for the ground electronic state in the nonrelativistic approximation, all digits are significant.}
 \label{NAPTresultsHD}
 \begin{ruledtabular}
 \begin{tabular}{x{1.1} x{1.8} x{1.8}}
\cent{R} & \cent{\delta \sigma(\mathrm{HD})} & \cent{\delta \sigma(\mathrm{HT})} \\ 
\hline\rule{0pt}{2.5ex}
0.1 & 0.051\,718\,4 & 0.076\,459\,4 \\
0.2 & 0.046\,110\,4 & 0.061\,866\,2 \\
0.3 & 0.040\,891\,0 & 0.052\,832\,8 \\
0.4 & 0.036\,764\,2 & 0.046\,485\,4 \\
0.5 & 0.033\,484\,3 & 0.041\,729\,0 \\
0.6 & 0.030\,864\,4 & 0.038\,063\,0 \\
0.8 & 0.026\,956\,0 & 0.032\,773\,7 \\
1.0 & 0.024\,208\,7 & 0.029\,168\,6 \\
1.1 & 0.023\,124\,0 & 0.027\,769\,0 \\
1.2 & 0.022\,182\,5 & 0.026\,564\,5 \\
1.3 & 0.021\,357\,3 & 0.025\,516\,3 \\
1.4 & 0.020\,627\,9 & 0.024\,595\,4 \\
1.5 & 0.019\,976\,4 & 0.023\,776\,8 \\
1.6 & 0.019\,391\,2 & 0.023\,044\,2 \\
1.7 & 0.018\,860\,3 & 0.022\,381\,2 \\
1.8 & 0.018\,374\,6 & 0.021\,775\,6 \\
1.9 & 0.017\,926\,3 & 0.021\,216\,5 \\
2.0 & 0.017\,507\,8 & 0.020\,693\,9 \\
2.1 & 0.017\,115\,0 & 0.020\,202\,2 \\
2.2 & 0.016\,741\,8 & 0.019\,733\,0 \\
2.3 & 0.016\,385\,1 & 0.019\,282\,5 \\
2.4 & 0.016\,041\,3 & 0.018\,845\,6 \\
2.5 & 0.015\,707\,4 & 0.018\,418\,3 \\
2.6 & 0.015\,381\,8 & 0.017\,998\,6 \\
2.7 & 0.015\,062\,8 & 0.017\,583\,8 \\
2.8 & 0.014\,750\,4 & 0.017\,174\,1 \\
2.9 & 0.014\,444\,4 & 0.016\,769\,0 \\
3.0 & 0.014\,145\,5 & 0.016\,369\,6 \\
3.2 & 0.013\,572\,0 & 0.015\,592\,1 \\
3.4 & 0.013\,041\,4 & 0.014\,857\,7 \\
3.6 & 0.012\,567\,9 & 0.014\,187\,7 \\
3.8 & 0.012\,160\,1 & 0.013\,596\,7 \\
4.0 & 0.011\,822\,0 & 0.013\,094\,8 \\
4.2 & 0.011\,549\,7 & 0.012\,680\,1 \\
4.4 & 0.011\,338\,2 & 0.012\,349\,5 \\
4.6 & 0.011\,177\,3 & 0.012\,090\,8 \\
4.8 & 0.011\,056\,7 & 0.011\,892\,3 \\
5.0 & 0.010\,968\,1 & 0.011\,742\,6 \\
5.2 & 0.010\,903\,4 & 0.011\,630\,4 \\
5.4 & 0.010\,856\,9 & 0.011\,547\,4 \\
5.6 & 0.010\,823\,3 & 0.011\,485\,8 \\
5.8 & 0.010\,799\,5 & 0.011\,441\,4 \\
6.0 & 0.010\,782\,8 & 0.011\,408\,9 \\
\end{tabular}
\end{ruledtabular}
\end{table}

\clearpage

\bibliographystyle{apsrev}

\begin{thebibliography}{20}
\expandafter\ifx\csname natexlab\endcsname\relax\def\natexlab#1{#1}\fi
\expandafter\ifx\csname bibnamefont\endcsname\relax
  \def\bibnamefont#1{#1}\fi
\expandafter\ifx\csname bibfnamefont\endcsname\relax
  \def\bibfnamefont#1{#1}\fi
\expandafter\ifx\csname citenamefont\endcsname\relax
  \def\citenamefont#1{#1}\fi
\expandafter\ifx\csname url\endcsname\relax
  \def\url#1{\texttt{#1}}\fi
\expandafter\ifx\csname urlprefix\endcsname\relax\def\urlprefix{URL }\fi
\providecommand{\bibinfo}[2]{#2}
\providecommand{\eprint}[2][]{\url{#2}}

\bibitem[{\citenamefont{Schneider et~al.}(2017)\citenamefont{Schneider, Mooser,
  Bohman, Schön, Harrington, Higuchi, Nagahama, Sellner, Smorra, Blaum
  et~al.}}]{Schneider:17}
\bibinfo{author}{\bibfnamefont{G.}~\bibnamefont{Schneider}},
  \bibinfo{author}{\bibfnamefont{A.}~\bibnamefont{Mooser}},
  \bibinfo{author}{\bibfnamefont{M.}~\bibnamefont{Bohman}},
  \bibinfo{author}{\bibfnamefont{N.}~\bibnamefont{Schön}},
  \bibinfo{author}{\bibfnamefont{J.}~\bibnamefont{Harrington}},
  \bibinfo{author}{\bibfnamefont{T.}~\bibnamefont{Higuchi}},
  \bibinfo{author}{\bibfnamefont{H.}~\bibnamefont{Nagahama}},
  \bibinfo{author}{\bibfnamefont{S.}~\bibnamefont{Sellner}},
  \bibinfo{author}{\bibfnamefont{C.}~\bibnamefont{Smorra}},
  \bibinfo{author}{\bibfnamefont{K.}~\bibnamefont{Blaum}},
  \bibnamefont{et~al.}, \bibinfo{journal}{Science}
  \textbf{\bibinfo{volume}{358}}, \bibinfo{pages}{1081} (\bibinfo{year}{2017}).

\bibitem[{\citenamefont{Wehrli et~al.}(2021)\citenamefont{Wehrli,
  Spyszkiewicz-Kaczmarek, Puchalski, and Pachucki}}]{Wehrli:21}
\bibinfo{author}{\bibfnamefont{D.}~\bibnamefont{Wehrli}},
  \bibinfo{author}{\bibfnamefont{A.}~\bibnamefont{Spyszkiewicz-Kaczmarek}},
  \bibinfo{author}{\bibfnamefont{M.}~\bibnamefont{Puchalski}},
  \bibnamefont{and} \bibinfo{author}{\bibfnamefont{K.}~\bibnamefont{Pachucki}},
  \bibinfo{journal}{Phys. Rev. Lett.} \textbf{\bibinfo{volume}{127}},
  \bibinfo{pages}{263001} (\bibinfo{year}{2021}).

\bibitem[{\citenamefont{Schneider et~al.}(2019)\citenamefont{Schneider, Mooser,
  Rischka, Blaum, Ulmer, and Walz}}]{schneider19}
\bibinfo{author}{\bibfnamefont{A.}~\bibnamefont{Schneider}},
  \bibinfo{author}{\bibfnamefont{A.}~\bibnamefont{Mooser}},
  \bibinfo{author}{\bibfnamefont{A.}~\bibnamefont{Rischka}},
  \bibinfo{author}{\bibfnamefont{K.}~\bibnamefont{Blaum}},
  \bibinfo{author}{\bibfnamefont{S.}~\bibnamefont{Ulmer}}, \bibnamefont{and}
  \bibinfo{author}{\bibfnamefont{J.}~\bibnamefont{Walz}},
  \bibinfo{journal}{Ann. Phys.} \textbf{\bibinfo{volume}{531}},
  \bibinfo{pages}{1800485} (\bibinfo{year}{2019}).

\bibitem[{\citenamefont{Garbacz
  et~al.}(2012{\natexlab{a}})\citenamefont{Garbacz, Jackowski, Makulski, and
  Wasylishen}}]{Garbacz:12}
\bibinfo{author}{\bibfnamefont{P.}~\bibnamefont{Garbacz}},
  \bibinfo{author}{\bibfnamefont{K.}~\bibnamefont{Jackowski}},
  \bibinfo{author}{\bibfnamefont{W.}~\bibnamefont{Makulski}}, \bibnamefont{and}
  \bibinfo{author}{\bibfnamefont{R.~E.} \bibnamefont{Wasylishen}},
  \bibinfo{journal}{J. Phys. Chem. A} \textbf{\bibinfo{volume}{116}},
  \bibinfo{pages}{11896} (\bibinfo{year}{2012}{\natexlab{a}}).

\bibitem[{\citenamefont{Ramsey}(1950)}]{Ramsey:50}
\bibinfo{author}{\bibfnamefont{N.~F.} \bibnamefont{Ramsey}},
  \bibinfo{journal}{Phys. Rev.} \textbf{\bibinfo{volume}{78}},
  \bibinfo{pages}{699} (\bibinfo{year}{1950}).

\bibitem[{\citenamefont{Neronov and Barzakh}(1977)}]{Neronov:77}
\bibinfo{author}{\bibfnamefont{Y.~I.} \bibnamefont{Neronov}} \bibnamefont{and}
  \bibinfo{author}{\bibfnamefont{A.~E.} \bibnamefont{Barzakh}},
  \bibinfo{journal}{Zh. Eksp. Teor. Fiz.} \textbf{\bibinfo{volume}{72}},
  \bibinfo{pages}{1659} (\bibinfo{year}{1977}).

\bibitem[{\citenamefont{Jaszu{\'{n}}ski
  et~al.}(2011)\citenamefont{Jaszu{\'{n}}ski, {\L}ach, and
  Strasburger}}]{Jaszunski:11}
\bibinfo{author}{\bibfnamefont{M.}~\bibnamefont{Jaszu{\'{n}}ski}},
  \bibinfo{author}{\bibfnamefont{G.}~\bibnamefont{{\L}ach}}, \bibnamefont{and}
  \bibinfo{author}{\bibfnamefont{K.}~\bibnamefont{Strasburger}},
  \bibinfo{journal}{Theor. Chem. Acc.} \textbf{\bibinfo{volume}{129}},
  \bibinfo{pages}{325} (\bibinfo{year}{2011}).

\bibitem[{\citenamefont{Golubev and Shchepkin}(2014)}]{Golubev:14}
\bibinfo{author}{\bibfnamefont{N.~S.} \bibnamefont{Golubev}} \bibnamefont{and}
  \bibinfo{author}{\bibfnamefont{D.~N.} \bibnamefont{Shchepkin}},
  \bibinfo{journal}{Chem. Phys. Lett.} \textbf{\bibinfo{volume}{591}},
  \bibinfo{pages}{292} (\bibinfo{year}{2014}).

\bibitem[{\citenamefont{Puchalski et~al.}(2015)\citenamefont{Puchalski, Komasa,
  and Pachucki}}]{Puchalski:15}
\bibinfo{author}{\bibfnamefont{M.}~\bibnamefont{Puchalski}},
  \bibinfo{author}{\bibfnamefont{J.}~\bibnamefont{Komasa}}, \bibnamefont{and}
  \bibinfo{author}{\bibfnamefont{K.}~\bibnamefont{Pachucki}},
  \bibinfo{journal}{Phys. Rev. A} \textbf{\bibinfo{volume}{92}},
  \bibinfo{pages}{020501} (\bibinfo{year}{2015}).

\bibitem[{\citenamefont{Neronov and Karshenboim}(2003)}]{Neronov:03}
\bibinfo{author}{\bibfnamefont{Y.~I.} \bibnamefont{Neronov}} \bibnamefont{and}
  \bibinfo{author}{\bibfnamefont{S.~G.} \bibnamefont{Karshenboim}},
  \bibinfo{journal}{Phys. Lett. A} \textbf{\bibinfo{volume}{318}},
  \bibinfo{pages}{126} (\bibinfo{year}{2003}).

\bibitem[{\citenamefont{Neronov and Aleksandrov}(2011)}]{Neronov:11}
\bibinfo{author}{\bibfnamefont{Y.~I.} \bibnamefont{Neronov}} \bibnamefont{and}
  \bibinfo{author}{\bibfnamefont{V.~S.} \bibnamefont{Aleksandrov}},
  \bibinfo{journal}{JETP Lett.} \textbf{\bibinfo{volume}{94}},
  \bibinfo{pages}{418} (\bibinfo{year}{2011}).

\bibitem[{\citenamefont{Pachucki and Komasa}(2008)}]{PK:08}
\bibinfo{author}{\bibfnamefont{K.}~\bibnamefont{Pachucki}} \bibnamefont{and}
  \bibinfo{author}{\bibfnamefont{J.}~\bibnamefont{Komasa}},
  \bibinfo{journal}{J. Chem. Phys.} \textbf{\bibinfo{volume}{129}},
  \bibinfo{eid}{034102} (pages~\bibinfo{numpages}{7}) (\bibinfo{year}{2008}).

\bibitem[{\citenamefont{Pachucki and Komasa}(2009)}]{PK:09}
\bibinfo{author}{\bibfnamefont{K.}~\bibnamefont{Pachucki}} \bibnamefont{and}
  \bibinfo{author}{\bibfnamefont{J.}~\bibnamefont{Komasa}},
  \bibinfo{journal}{J. Chem. Phys.} \textbf{\bibinfo{volume}{130}},
  \bibinfo{eid}{164113} (pages~\bibinfo{numpages}{11}) (\bibinfo{year}{2009}).

\bibitem[{\citenamefont{Komasa et~al.}(2019)\citenamefont{Komasa, Puchalski,
  Czachorowski, \L{}ach, and Pachucki}}]{Komasa:19}
\bibinfo{author}{\bibfnamefont{J.}~\bibnamefont{Komasa}},
  \bibinfo{author}{\bibfnamefont{M.}~\bibnamefont{Puchalski}},
  \bibinfo{author}{\bibfnamefont{P.}~\bibnamefont{Czachorowski}},
  \bibinfo{author}{\bibfnamefont{G.}~\bibnamefont{\L{}ach}}, \bibnamefont{and}
  \bibinfo{author}{\bibfnamefont{K.}~\bibnamefont{Pachucki}},
  \bibinfo{journal}{Phys. Rev. A} \textbf{\bibinfo{volume}{100}},
  \bibinfo{pages}{032519} (\bibinfo{year}{2019}).

\bibitem[{\citenamefont{Tiesinga et~al.}(2021)\citenamefont{Tiesinga, Mohr,
  Newell, and Taylor}}]{CODATA:18}
\bibinfo{author}{\bibfnamefont{E.}~\bibnamefont{Tiesinga}},
  \bibinfo{author}{\bibfnamefont{P.~J.} \bibnamefont{Mohr}},
  \bibinfo{author}{\bibfnamefont{D.~B.} \bibnamefont{Newell}},
  \bibnamefont{and} \bibinfo{author}{\bibfnamefont{B.~N.}
  \bibnamefont{Taylor}}, \bibinfo{journal}{Rev. Mod. Phys.}
  \textbf{\bibinfo{volume}{93}}, \bibinfo{pages}{025010}
  (\bibinfo{year}{2021}).

\bibitem[{\citenamefont{Pachucki}(2010)}]{Pachucki:10}
\bibinfo{author}{\bibfnamefont{K.}~\bibnamefont{Pachucki}},
  \bibinfo{journal}{Phys. Rev. A} \textbf{\bibinfo{volume}{81}},
  \bibinfo{pages}{032505} (\bibinfo{year}{2010}).

\bibitem[{\citenamefont{Puchalski et~al.}(2018)\citenamefont{Puchalski,
  Spyszkiewicz, Komasa, and Pachucki}}]{Puchalski:18}
\bibinfo{author}{\bibfnamefont{M.}~\bibnamefont{Puchalski}},
  \bibinfo{author}{\bibfnamefont{A.}~\bibnamefont{Spyszkiewicz}},
  \bibinfo{author}{\bibfnamefont{J.}~\bibnamefont{Komasa}}, \bibnamefont{and}
  \bibinfo{author}{\bibfnamefont{K.}~\bibnamefont{Pachucki}},
  \bibinfo{journal}{Phys. Rev. Lett.} \textbf{\bibinfo{volume}{121}},
  \bibinfo{pages}{073001} (\bibinfo{year}{2018}).

\bibitem[{\citenamefont{Puchalski et~al.}(2019)\citenamefont{Puchalski, Komasa,
  Spyszkiewicz, and Pachucki}}]{Puchalski:19}
\bibinfo{author}{\bibfnamefont{M.}~\bibnamefont{Puchalski}},
  \bibinfo{author}{\bibfnamefont{J.}~\bibnamefont{Komasa}},
  \bibinfo{author}{\bibfnamefont{A.}~\bibnamefont{Spyszkiewicz}},
  \bibnamefont{and} \bibinfo{author}{\bibfnamefont{K.}~\bibnamefont{Pachucki}},
  \bibinfo{journal}{Phys. Rev. A} \textbf{\bibinfo{volume}{100}},
  \bibinfo{pages}{020503} (\bibinfo{year}{2019}).

\bibitem[{\citenamefont{Neronov and Seregin}(2012)}]{Neronov:12}
\bibinfo{author}{\bibfnamefont{Y.~I.} \bibnamefont{Neronov}} \bibnamefont{and}
  \bibinfo{author}{\bibfnamefont{N.~N.} \bibnamefont{Seregin}},
  \bibinfo{journal}{Journal of Experimental and Theoretical Physics}
  \textbf{\bibinfo{volume}{115}}, \bibinfo{pages}{777} (\bibinfo{year}{2012}).

\end{thebibliography}

\end{document}